\documentclass[onecolumn,showkeys,showpacs,superscriptaddress,notitlepage]{revtex4-1}
\usepackage{graphicx}
\usepackage{amsmath,amsfonts,amssymb}
\usepackage{morefloats}
\usepackage{color}
\usepackage{subcaption}
\usepackage{multirow}

\newcommand{\be}{\begin{equation}}
\newcommand{\ee}{\end{equation}}

\begin{document}
\title[ ]{Self-presentation and emotional contagion on Facebook: new experimental measures of profiles' emotional coherence}

\author{Andrea Guazzini}\email{andrea.guazzini@unifi.it}
\affiliation{Department of Education and Psychology, and Center for the Study of Complex Dynamics (CSDC), University of Florence, Via di San Salvi 12, 50135, Florence, Italy}
\author{Elisa Guidi}\email{elisa.guidi@unifi.it}
\affiliation{Department of Information Engineering, via S. Marta 3, I-50193, Florence (Italy)}
\author{Cristina Cecchini}\email{crisitna.cecchini@unifi.it}
\affiliation{Department of Information Engineering, via S. Marta 3, I-50193, Florence (Italy)}
\author{Monica Milani}\email{monica.milani@unifi.it}
\affiliation{Department of Information Engineering, via S. Marta 3, I-50193, Florence (Italy)}
\author{Daniele Vilone}\email{daniele.vilone@gmail.com}
\affiliation{LABSS (Laboratory of Agent Based Social Simulation),
Institute of Cognitive Science and Technology,
National Research Council (CNR), 
Via Palestro 32, 00185 Rome, Italy}
\affiliation{Grupo Interdisciplinar de Sistemas Complejos (GISC), Departamento de Matem\'aticas,
Universidad Carlos III de Madrid, 28911 Legan\'es (Spain)}
\author{Patrizia Meringolo}\email{patrizia.meringolo@unifi.it}
\affiliation{Department of Education and Psychology, and Center for the Study of Complex Dynamics (CSDC), University of Florence, Via di San Salvi 12, 50135, Florence, Italy}

\maketitle

\section*{abstract}
Social Networks allow users to self-present by sharing personal contents with others which may add comments. Recent studies highlighted how the emotions expressed in a post affect others' posts, eliciting a congruent emotion. So far, no studies have yet investigated the emotional coherence between wall posts and its comments. This research evaluated posts and comments mood of Facebook profiles, analyzing their linguistic features, and a measure to assess an excessive self-presentation was introduced. Two new experimental measures were built, describing the emotional loading (positive and negative) of posts and comments, and the mood correspondence between them was evaluated. The profiles "empathy", the mood coherence between post and comments, was used to investigate the relation between an excessive self-presentation and the emotional coherence of a profile. Participants publish a higher average number of posts with positive mood. To publish an emotional post corresponds to get more likes, comments and receive a coherent mood of comments, confirming the emotional contagion effect reported in literature. Finally, the more empathetic profiles are characterized by an excessive self-presentation, having more posts, and receiving more comments and likes. To publish emotional contents appears to be functional to receive more comments and likes, fulfilling needs of attention-seeking.

%\begin{keyword}
%Social Networks \sep Adolescents \sep Virtual Dynamics \sep Self-promotion \sep Social Psychology \sep Computer-mediated communication
%\end{keyword}

\section{Introduction}
\label{intro}
The rise of World Wide Web offered new tools to spread news and information, managing any time. The direct outcome was to increase interactions among different Internet users \cite{Bohna2014}, preparing a suitable ground for the birth of Social Networks (SNSs). SNSs are an interesting growing field \cite{Goodings2007,Whittaker2013} for Psychology, Computer Science and Sociology, because their popularity allows researchers to open up new opportunities for reliable affective assessment \cite{Kanjo2015,VandenBroek2013}. At present Facebook is one of most used SNSs \cite{Day2013}, and in last decades it has developed an innovative way to create and increase social relationships of people of all ages \cite{Bazarova2012,Niland2015}. Particularly, these technologies let users sharing personal contents on their profile, allowing others to read and add comments \cite{O'Sullivan2005}. This is basic for online self-presentation, where users need to find effective strategies to fully express themselves with different audiences \cite{Kramer2011,Lampe2010,Papacharissi2011}. Facebook users easily change their approach and communication with others \cite{Richardson2009}, which may indicate they apply different strategies depending on kind of audiences and type of published message (i.e., public wall posts, wall posts updates, private messages). Choosing what to discover and which emotion may produce positive or negative reactions from audience \cite{Bazarova2012}. Some studies \cite{Leary1995,Leary1990} state users seek two main goals through self-presentation: impression construction, to create a desired impression, and impression motivation, to manage others' opinion of the self.  This is complicated on SNSs, where audience is various and rather difficult to select \cite{Kramer2011}. Then, the best strategy appears to be to show only acceptable information for all targets \cite{Hogan2010}, while superficial messages appear not to be suitable in creating positive impressions \cite{Bazarova2012}. However, such information represents a self-disclosure of personal issues, such as attitudes, personal hobbies, experiences and emotions. A recent research \cite{Bazarova2012} investigated  how the use of emotions, both positive and negative, affected the linguistic style for self-presentation on Facebook. Findings confirmed different adoptions of language style depending on the audience, and highlighted a major use of positive emotions, as opposed to negative emotions, in status updates. Therefore, emotions appear to be central not only in in-person interactions, but also in online communication \cite{Garas2012,Hu2015}, where they can be spread and reach other people, who may be affected in a sort of "empathetic contagion" \cite{Dimberg2012}. Indeed, one relevant question was whether this contagion would be effective in the cyber-world. Research exploring online empathy has especially examined support communities or "thematic" forums (e.g., sports) \cite{Preece2001,Pfeil2007}, or Instant Messaging \cite{Feng2004}, analyzing empathetic responses to the posts. In these studies, empathy was defined as the capacity to both feel and understand others' feelings and thoughts \cite{Lawrence2004}, which allow to predict others' intentions and experience the same emotion \cite{Baron2004,Eisenberg1997}. Empathy is basic to build appropriate social relationships \cite{Venkatanathan2013} and is a predictor of pro-social behaviors \cite{Eisenberg2000,Miller1997} and perspective taking abilities \cite{Findlay2006}. Studies on online empathy revealed that online support in thematic communities consisted of providing practical information and fostering emotional attachment \cite{Wright2000}. Participating members build social relationships and are more likely to write empathetic messages than new members \cite{Pfeil2007}. Moreover, \cite{Feng2004} found that an empathetic communication had a significant influence on online interpersonal trust, where more empathetic people were more trusted by others.  Moreover, a study enhancing gender differences in virtual communities underlined how females published more empathetic messages than males \cite{Preece2001}. Facial and body expressions can be helpful to empathy in understanding one's behavior \cite{Batson2014}, but online interactions are missing the use of non-verbal communication, complicating the empathetic answers \cite{Culnan1987}. By contrast, the virtual environments are feasible to precisely detect and record every gesture, voice feature, non verbal behavior, avoiding difficult data mining/coding of empathy and emotional contagion in real settings.
We recently tested such aspects, confirming that the mood of a short message, such as those observed within online social networks, forums, and web-based chats, can be detected and be informative about the dynamics of the system and the topological position of the writer \cite{Cini2013}.
The concept of "virtual empathy” (i.e., the existence of an empathy capacity for humans into virtual environments) could state or predict a measurable effect of the mood polarization of a stimulus (e.g., messages, photos) on the inner psychological state of the observer. As a consequence, some effects on the subsequent production of the observer should be detectable, as well as a possible coupling between the behavior of the observer and the post maker.
Regarding this, some studies began to explore the computer-mediated communication \cite{Alberici2013} and the spread of emotions \cite{Hancock2008,Guillory2011}. Results showed that inducing negative emotions elicited negative messages in participants, confirming that the emotional contagion on virtual environments is not only detectable, but clearly evident even without face-to-face interactions. Recently, emotional dissemination was analyzed on Facebook, examining posts and status updates. Findings showed emotions expression conditioned others' emotional posts, eliciting a congruent emotional contagion: that is, people having friends who published positive posts were more likely to publish positive messages as well. Then, emotions transmission does not occur only after in-person interactions, but also through computer-mediated communication \cite{Kramer2012,Kramer2014}.
Despite such interesting findings, no studies have yet investigated emotional coherence between wall posts and received comments. When a user publishes a status, are comments he receives emotionally congruent with that status? The present research purposes to explore this aspect by means of the development of two experimental metrics assessing the emotional loading and the emotional coherence.

\section{Excessive Self-presentation, Narcissistic Trait and Emotional Mood of Posts}
\label{sec:1}

Self presentation on SNSs has been largely explored in literature, and researchers also investigated the origins of an excessive online self-presentation, wondering whether a frequent posting might be a signal of a narcissistic trait \cite{Bergman2011,Carpenter2012,Deters2014,Rosen2013}. Literature described narcisissm \cite{Campbell2007,Mehdizadeh2010,Vazire2008} as a personality trait and not a clinical disorder, and as the tendency to an inflated and positive self-concept, with exhibitionism and attention-seeking behaviors.
SNSs are suitable to achieve narcissistic goals, and they may also incentive to self-promote and to engage superficial behaviors \cite{Buffardi2008}. \cite{Ryan2011} confirmed this, finding Facebook users to be more narcissistic than non-users.
Other studies referred that narcissistic and excessive self-presentation strategies on SNSs were related to publish more self-promotional information \cite{Panek2013} and to have more friends \cite{Bergman2011,Buffardi2008}, besides using more swear and anger words \cite{Carpenter2012}, sexual words \cite{Holtzman2010} and singular first-person pronouns \cite{DeWall2011}. It appears that narcissistic people engage in more self-promotional and self-disclosed behaviors, as frequent status updates, to seek for others' attention  \cite{Ong2011}. A study investigated how personality traits influenced self-presentation, self-disclosure and linguistic and emotional content of messages on Facebook \cite{Winter2014}. Results displayed narcissists disclose more personal information and self-present more than others, revealing a tendency to an excessive self-presentation. Despite a frequent negative relation between narcissism and empathy on SNSs and the widespread knowledge about SNSs promoting narcissism \cite{Buffardi2008}, another research revealed that participating to SNSs has an association with empathy, too \cite{Gountas2011}. To investigate the presence of emotional loading in SNSs narcissistic posts, and the association between online narcissism and empathy, is nowadays of particular interest because of the recent findings about the negative correlation between empathetic behaviors and the narcissistic style \cite{Konrath2010}. Moreover, more information about linguistic strategies in the narcissistic trait and in the excessive self-presentation could be provided. Unfortunately, few studies analyzed these aspects. Given that some studies explored linguistic contents in the  narcissistic trait and excessive self-presentation on Facebook \cite{DeWall2011,Holtzman2010}, revealing a peculiar use of communication strategies for attention-seeking goals \cite{Buss1991}, a second purpose of the present study was to analyze all posts of 50 Facebook profiles in one year through the linguistic software LIWC \cite{Pennebaker2007}, and to explore content emotional features of posts and comments in excessive self-presentation profiles. Particularly, we meant to identify an "empathetic coherence or incoherence" between posts and received comments in different profiles.

\section{Aims of our study}
\label{sec:1}

The main challenge of our study concerns about the dynamics (e.g., spreading and sensitiveness) of the emotional coherence on Facebook among Italian adolescents. We focused on adolescents' sample because recent studies underlined how most of members of SNSs are young people \cite{Jordan2014}. The first property required to the measure is the ability to detect the emotional content (i.e., sentiment) of a web-based post (e.g., message, photo, news, etc). Given the literature about the "Sentiment Analysis” \cite{Hu2013,Hu2013exploiting,Tan2011}, several tools and approaches can be adopted to fulfill this first challenge. For this reason, our study analyzed $50$ Facebook users' profiles, coding each published post or activity during a year (table \ref{tab:FacebookFeatures}). In our study we first define for all the Facebook profile an emotional loading for each post and comment by means of the development of two operative metrics (i.e., negative and positive mood indicators). Once the emotional loading of each post and comment was evaluated, the “emotional coherence” of each profile has been defined as the normalized correspondence between each post mood, and its average comments moods, assessed by the Pearson $\chi^2$ statistics. As a consequence, the empathy level of a profile is defined as the degree of agreement between the moods of the posts, and the moods of the comments received by each post. Finally, the average emotional loading and coherence of profiles have been related with the gender, with the variables describing the social network usage, and with the self-presentation style of individuals.

\section{Methods}
\label{Methods}

\subsection{Participants and procedure}
\label{sec:2}
Participants were $50$ students ($50\%$ females) recruited from a Tuscan high school. They ranged in age from $15$ to $19$ (M=$16.95$, SD=$1.08$). All participants had a Facebook account and were involved as volunteers.
The data collection carried out during the “ARCA project". The agreement of the high school to participate was obtained from the principal. The professors of the classes involved, and a research assistant introduced the aims of the survey, and the confidentiality issues to the students. Before the students' participation in the research, parental consent and adolescent assent was obtained. 
To analyze one year of the Facebook usage on participants' profiles, a Facebook account was created using the recruitment coordinator's contact information, with a research logo as the profile picture. Before befriending a participant's profile, we sent a private message inviting him/her to participate in the research. Participants were explicitly notified that the researchers would have access to their Facebook profiles for one year and we communicated that the data would be stored anonymously. All the 50 students gave permission to save their Facebook pages to be used in the present research.

\begin{table*}[htbp!]
\caption[Facebook Profile Features]{\textit{Facebook Profile Features}. The descriptive statistics related to the sample's Facebook profiles features are reported}
\label{tab:FacebookFeatures}
\begin{center}
\small
\begin{tabular}{lcccc}
\multicolumn{1}{c}{\bf Facebook Variable}  &\multicolumn{1}{c}{\bf M} &\multicolumn{1}{c}{\bf SD} &\multicolumn{1}{c}{\bf Min} &\multicolumn{1}{c}{\bf Max}
\\ \hline \\
Friends & 3085 & 1089 & 642 & 4970\\
Followed people & 12 & 11 & 0 & 46\\
Visited places & 64 & 124 & 1 & 644\\
Famous quotes & 94 & 99 & 1 & 333\\
Pages with likes & 547 & 781 & 30 & 4762\\
Complete activity & 647 & 529 & 29 & 2609\\
Wall posts & 578 & 491 & 22 & 2399\\
Profile picture edit & 23 & 20 & 0 & 81\\
Personal photos & 49 & 43 & 1 & 206\\
Photos & 218 & 194 & 15 & 908\\
Videos & 62 & 93 & 1 & 562\\
Likes & 12646 & 13280 & 465 & 66815\\
Activities with like & 641 & 520 & 29 & 2537\\
Wall post with comments & 263 & 179 & 21 & 746\\
Comments & 1242 & 926 & 121 & 4494\\
Wall posts length & 46814 & 42021 & 2010 & 193966\\
Wall posts average length & 81 & 30 & 31 & 183\\
\end{tabular}
\normalsize
\end{center}
\end{table*}

\subsection{Measures}
\label{sec:2} 

\paragraph{Facebook Page Coding (observation grid)}
From each Facebook profile, some relevant information were extracted, both concerning the  directly available data (i.e., friends, followed people, visited places, famous quotes, pages with like), and some objective criteria calculated and coded by the year-long analysis of each profile (i.e. complete activity, wall posts, profile picture edit, personal photos, photos, videos, likes, activities with likes, comments, posts with comments, wall posts length, wall posts average length). 
We analyzed one year of participants' activity on Facebook, considering a total of $32368$ activities($28878$ of which were wall posts), and $62083$ comments.

\paragraph{Linguistic Analysis} 

In order to assess the posts emotional loading, we used the Linguistic Inquiry Word Count program \cite{Pennebaker2007}. LIWC analyzes transcripts on a word-by-word basis and compares words with a dictionary related to $70$ linguistic, affective, cognitive and social processes. LIWC’s psychometric properties and external validity have been established in a large number of studies, and has been used to examine the relationship between language and emotion, personality, and deception, among others \cite{Tausczik2010}.
In this study, all the sampled profiles have been considered as separated elements, and the LIWC analysis concerned the entire production (i.e., the posts) as a single narration (table \ref{tab:5-Anova}). 

\begin{table*}[htbp!]
\caption[Anova]{In table are reported the LIWC categories discriminating effectively the posts/comments mood. In particular the significant Fischer F, and the associated sum of squares are reported and connected with the specific mood assessed by them. The signs between brackets after the mood polarization indicate the sign assumed by the terms in the operative models.}
\label{tab:5-Anova}
\begin{center}
\small
\begin{tabular}{lccc}
\multicolumn{1}{c}{\bf Variable}  &\multicolumn{1}{c}{\bf Sum of Squares} &\multicolumn{1}{c}{\bf F($\textit{p.}<0.01$)} &\multicolumn{1}{c}{\bf Condition}
\\ \hline \\
Negative emotions & 4.009 & 21,063 & Negative [+]\\
Swear word & 5.356 & 12,418 & Negative [+]\\
Positive feeling & 2.002 & 12,309 & Positive [+]\\
Anger word & 6.000 & 11,355 & Negative [+]\\
Positive Emotion & 2.839 & 10,169 & Positive [+]\\
Sadness word & 4.562 & 9,834 & Negative [+]\\
Numerals & 0.939 & 8,376 & Negative [-]\\
Third-Person plural verb & 3.441 & 6,214 & Negative [+]\\
Family & 5.631 & 5,937 & Positive [+]\\
Question marks & 3.374 & 5,587 & Positive [+]\\
\end{tabular}
\normalsize
\end{center}
\end{table*}

\paragraph{Excessive Self-presentation Model} 

Since excessive online self-presentation could be an indicator of narcissism \cite{Deters2014,Rosen2013}, we used a previously developed model, labeled “Excessive online self-presentation model”, to directly assess this tendency, through the linguistic analysis of public contents of personal pages. This measure was originated by an observation grid coding the Facebook activities of our participants (e.g. wall posts, comments, photos, etc.), and the analysis of the language style obtained through LIWC analysis of their Facebook published posts. For more details please refer to \cite{Guazzinixx}. Such a measure has been defined investigating the linguistic features of the individuals' Facebook wall posts, merging recent studies validating online measurements of the narcissistic trait \cite{Bergman2011,Buffardi2008,Carpenter2012,DeWall2011,Holtzman2010,Panek2013}, with the theoretical models coming from classical literature. The final model presented in \cite{Guazzinixx} is composed by the two LIWC-based parameters (i.e. \textit{“Word Count” and “Sexual”}) provided by the linear regression analysis.

\paragraph{Creating metrics to measure emotional load and emotional coherence}

The assessment of \textit{Profiles Emotional Coherence} (i.e., Facebook Empathy Profile) includes a Sentiment Analysis of the wall posts and the comments to identify their "mood". In particular, the Facebook Empathy profile level is not the user's level, but it is the commentators' average level, in other words the comments accord level with a particular wall post. It was possible to define an Emotional Load of the wall post and its comments and, subsequently, a \textit{Profiles Emotional Coherence} defined as the degree of correspondence normalized between the wall posts mood and its comments. 
The Sentiment Analysis was conducted starting from the LIWC software analysis. Based on the LIWC categories we defined two metrics to estimate the emotional content of posts and comments: the Positive Mood Indicator and the Negative Mood Indicator.

\subsection{Data Analysis}
\label{sec:2}

\paragraph{Emotional coherence and loading indicators development}

\begin{figure*}[htbp!]
\begin{center}
  \includegraphics[width=0.7\textwidth]{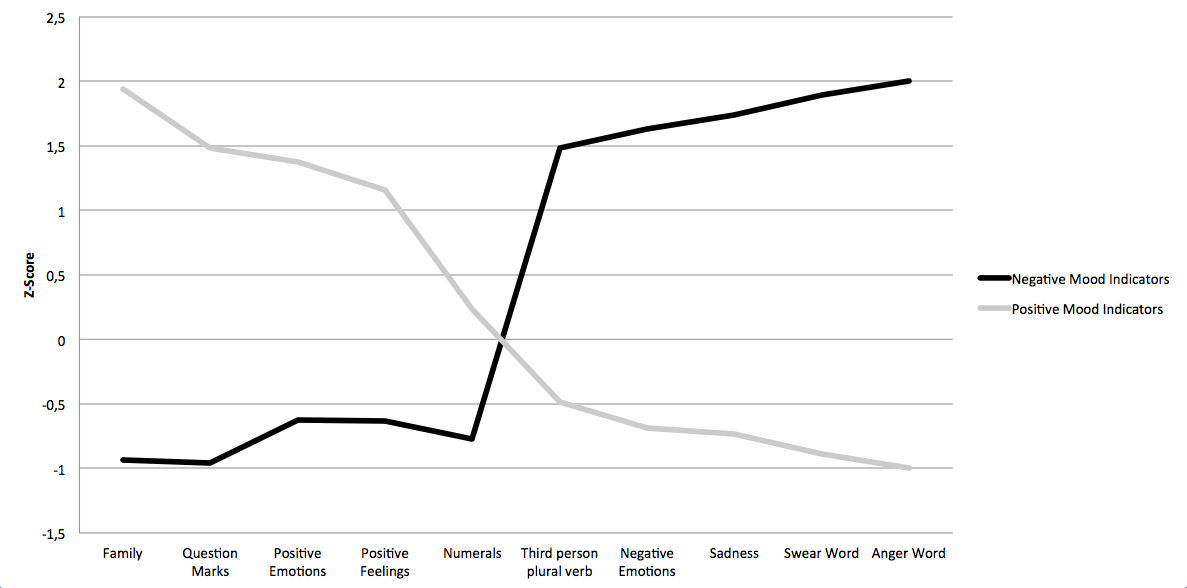}
\caption{\label{fig:Figure1} The z-scores associated with the LIWC variables significantly related with the operative criteria, and concerning the mood polarization are reported. }
\end{center}
\end{figure*}

Step $1$: The wall posts that did not receive any comments and those that were photos, videos, music, or famous quotes were excluded from the original sample ($14644$). Four judges examined a random sample of $500$ wall posts extracted from the new set of $14234$.  The judges selected $144$ wall posts, in order to achieve $48$ wall posts defined by them as "Negative", $48$ as "Positive" and $48$ as "Neutral", comparing their emotional content. These subgroups of wall posts were used as a criterion to develop the model for the sentiment analysis.  
Step $2$: It was carried out an analysis of variance (ANOVA) to identify which LIWC variables showed a discerning capacity in identifying the condition of "Negative Mood Indicator", "Positive Mood Indicator", or "Neutral Mood Indicator". We chose the LIWC categories that reported a significant F Fischer's score. Moreover, to assess the main effects and to evaluate which condition the LIWC variable was able to discriminate, we  adopted the Scheff\' e test (Scheff\'e, 1999), and the most discriminating variables were organized and ordered by F-test score (Table 1). A greater value of F-score corresponded to a greater ability to discriminate an emotional condition with respect to the other two (e.g., positive mood against negative and neutral mood). 
Step $3$: After identifying the LIWC most discriminating categories, we have also calculated the Z-scores (i.e., associated to the F values) of such categories for each condition "Negative Mood Indicator",  "Positive Mood Indicator", or  "Neutral Mood Indicator". After sorting the Z scores (figure \ref{fig:Figure1}), the LIWC variables that reported higher average values on a specific mood and lower values on the other two were selected as the "best" predictors. 
Step $4$: Based on the LIWC Best predictors, we have defined three metrics to estimate the emotional content of posts and comments (Negative Mood Indicator, Positive Mood Indicator and Neutral Mood Indicator). After metrics verification, we decided to not consider the posts with neutral mood, despite the clarity construct emerging from the LIWC semantic and syntactic categories analysis. This decision was motivated by the huge variability exhibited by neutral mood wall posts (as opposed to positive and negative mood wall posts), and thus to avoid the risk of excessive and impossible generalization of the neutrality construct. 

\paragraph{Inferential analysis}

The statistical analysis comprised five different steps. In the first one, the \textit{Student t} statistic has been used to compare the sub-samples of posts categorized as \textit{positive}, with those with a \textit{negative} emotional loading. The differences between the two sub-samples have been investigated on the observable quantities defined as order parameters of our study (i.e., gender, Facebook variables, Positive and Negative Mood Indicators of comments). To balance the sub-samples, a bootstrap sampling has been adopted and a final sample of $3392$ units has been analyzed. The same strategy has been used to compare the posts with an emotional loading different from $0$ (i.e., positive plus negative posts), and those with no emotional loading (i.e., neutral posts). In this case the two sub-samples resulting from the bootstrap method were composed by approximately $6380$ units. To estimate the Facebook Empathy profile level, we used the Pearson chi square test, assessing the agreement degree  between the emotional loading of each post and its comments. To evaluate the relation between the empathy and the excessive self-presentation, we simply computed the Pearson correlation between the scores reported to the Chi square test, and all the observable quantities of interest. Finally, to compare the participants with an \textit{High empathy profile} with those with a \textit{Low empathy profile}, the sample has been split in two, depending on the median reported to the Chi Square test ($Median= 0.74$). Subsequently, another independent sample Student \textit{t} test has been calculated with respect to all the independent variables.

\section{Results}
\label{Results}
\subsection{Creation of two metrics to weight the emotional load of a message}
\label{sec:2}

The analysis of variance (ANOVA) on the $144$ posts selected by judges (i.e., $48$ negative, $48$ positive and $48$ neutral) was focused on identifying which LIWC variables showed a discerning capacity in the three conditions and to calculate two metrics (Positive Mood Indicator and Negative Mood Indicator). A greater $F$ Fischer's score corresponds a greater capacity to discriminate one of the emotional conditions with respect to the other. 
Table $1$ shows the LIWC categories that have obtained a significant $F$ Fischer, sorted in descending order in respect to its score. Some significant categories, such as body, and comma, were not able to discriminate the emotional content of the posts (i.e., test Scheff\'e not significant or negligible effect) and therefore they were not used to build the two metrics.

Furthermore, we have calculated the Z-scores for the LIWC variables that were able to discriminate the post emotional content, i.e. to produce two different scores for each post respectively indicating the positive and negative "load" of the message. The resulting graph (figure \ref{fig:Figure1}) shows that there is a clear inverse correspondence between the negative mood variables and those of positive mood. As shown in equation \ref{eq:positive}, regarding Positive Mood Indicator, the LIWC best predictors were the variables Positive Feeling (PF), Positive Emotion (PE), Family (Fa), and Question Marks (QM), while concerning Negative Mood Indicator (eq. \ref{eq:negative}) the LIWC best predictors were the categories Negative Emotion (NE), Swear Word (SW), Anger Word (AW), Sadness Word (SaW), Numerals (Nu), and Third-Person Plural Verb (TP).

\begin{equation}
\label{eq:positive}
PM_i^N=NE+SW+AW+SaW-Nu+TP
\end{equation}

\begin{equation}
\label{eq:negative}
PM_i^P=PF+PE+Fa+QM
\end{equation}

where, $PM^P$ and $PM^N$ respectively indicate the positive and negative \textit{mood score} for the post $i$.

The several variables of the models were added or subtracted to the total scores if Z-score sign was positive or negative (e.g., the LIWC variable \textit{"Negative Emotion"} appeared as characterized by a Z-score of $1.63$ for what concerns the negative load, and a Z-score of $-0.69$ for the positive load; while on the other hand, the variable \textit{"Numerals"} obtains a negative load of $-0.77$, and a positive load of $0.23$). Therefore, in the former case the Negative Emotion variable has been adopted as a positive term to add for the\textit{ Negative Mood Indicator} (i.e., its presence increases the probability to have a post with a negative mood), while the latter Numerals variable has been adopted as negative term to subtract in the same model, because its presence decreases the probability to have a negative post.

The descriptive statistics related to the Facebook profiles' features are reported in table \ref{tab:FacebookFeatures}.
Regarding the $14234$ posts, it has been calculated the descriptive statistics relating to the variables number of likes, number of comments and scores of the negative and positive mood variables.  Each post had a mean score of $18.15$ (SD = $23.154$) for number of likes and $2.06$ (SD = $4.078$) for number of comments. The average score for the Negative Mood Indicator was $4.915$ (SD = $15.939$) and for the Positive Mood Indicator was $5.776$ (SD = $12.689$). 
Concerning the comments, it has been calculated the average score of the negative (M = $.877$, SD = $5.204$) and positive mood indicators, too (M = $3.046$, SD = $12.62$). The frequency distributions of  the mood indicator variables do not appear normal (i.e., skewness was large). This is due to the abundance of posts and comments with a zero score. These posts ($6380$) and comments ($9940$) have been removed from the analysis, since they are not emotionally determined. Thus, about the $14234$ posts, those deemed valid were $7854$ ($55.2\%$) of which $4462$ identified as positive and $3392$ negative, while comments were deemed valid $4294$ ($30.2\%$), of which $3147$ positive and $1147$ negative. 
\subsection{Descriptive analysis}
\label{sec:2}

A preliminary analysis tested that no gender differences are detectable for what concerns all the order parameters such as number of likes, number of comments, negative comments scores and positive comments scores. 
The summary of the inferential analysis are reported in table \ref{tab:emotional_posts_differences}.
First, it was carried out a balance through bootstrap method to conduct a Student's t-test in which the sub-sample of positive posts has been compared to that of the negative posts with respect to the variables mentioned before. Two groups were obtained characterized by a same high number of units ($3392$). From the analysis carried out, there was a unique significant relationship between mood posts and negative comments. The negative posts have a significantly higher probability of receiving a negative comment ($t = 3.523, \textit{p.} < .01$).
Subsequently, another Student's t-test was performed to compare posts expressing emotional content (negative or positive) and posts obtaining a zero score for the positive and negative mood variables, in other words the $6380$ posts that had previously been removed from the analysis. The sub-sample containing emotional posts ($6380$) was obtained using the bootstrap method. The analysis showed that the emotional posts received significantly more "Likes" ($t = -6.667, p. < .01$), more comments ($t = -2.8, p. < .01$), more negative comments ($t = -2.82, p. < .01$), compared to neutral posts.

\begin{table*}[htbp!]
\caption[]{\textit{Differences between Negative VS Positive posts, and Neutral Vs Emotional posts, with respect to the number of likes, number of comments, negative comment scores, positive comment scores.}. The table reports the significant Student \textit{t} values describing the differences between groups. All the statistics reported in table are significant at a level of $\textit{p.} < 0.01 (**)$.}
\label{tab:emotional_posts_differences}
\begin{center}
\small
\begin{tabular}{lcccc}
\multicolumn{1}{c}{\bf Variable} &\multicolumn{1}{c}{\bf \textit{t}} &\multicolumn{1}{c}{\bf Group} &\multicolumn{1}{c}{\bf M} &\multicolumn{1}{c}{\bf SD} \\
\\ \hline \\
\multirow{2}{*}{Number of likes} & \multirow{2}{*}{1.748} &
Negative Posts & 21.30 & 23.02\\
&& Positive Posts & 20.31 & 23.42\\
\multirow{2}{*}{Number of Comments} & \multirow{2}{*}{-0.021} &
Negative Posts & 2.09 & 4.12\\
&& Positive Posts & 2.09 & 3.70\\
\multirow{2}{*}{Negative comments scores} & \multirow{2}{*}{-3.523**} &
Negative Posts & 1.25 & 5.79\\
&& Positive Posts & 0.79 & 5.10\\
\multirow{2}{*}{Positive comments scores} & \multirow{2}{*}{-1.007} &
Negative Posts & 3.05 & 10.39\\
&& Positive Posts & 3.78 & 15.79\\
\\ \hline \\
\multirow{2}{*}{Number of likes} & \multirow{2}{*}{6.667**} &
Neutral Posts & 15.37 & 23.00\\
&& Emotional Posts & 17.95 & 20.39\\
\multirow{2}{*}{Number of Comments} & \multirow{2}{*}{-2.800**} &
Neutral Posts & 1.93 & 4.16\\
&& Emotional Posts & 2.13 & 3.96\\
\multirow{2}{*}{Negative comment scores} & \multirow{2}{*}{-2.820**} &
Neutral Posts & 0.76 & 4.96\\
&& Emotional Posts & 1.03 & 5.51\\
\multirow{2}{*}{Positive comment scores} & \multirow{2}{*}{-1.716} &
Neutral Posts & 2.87 & 12.52\\
&& Emotional Posts & 3.27 & 13.63\\
\\ \hline \\
\end{tabular}
\normalsize
\end{center}
\end{table*}

\subsection{Tendency to the emotional coherence between post and comments}

After classifying all the wall posts and comments in relation to the mood, we calculated the Chi-square statistic, both for the entire sample and for each subject separately, to assess the degree of consistency between the post mood and its comments mood. We measured both a general level of empathy profiles, and a particular level of empathy for each individual profile. 
The general Chi-square shows a significant relationship between post mood and comments mood ($Chi^2 = 26.44, \textit{p.} < .01$). There is a tendency to mainly respond in a positive way to positive than negative posts. Nevertheless, the two posts categories elicit the same absolute number of negative comments. Thus, positive posts show lesser percentage of negative comments than negative posts because they are greater.

\paragraph{Classification of the subjects on the basis of the Emotional Coherence of their Facebook profile: empathic detector}

We classified the individuals' profiles to discriminate those "more" empathetic, that is profiles for which the relationship between comments and posts appeared to be stronger (i.e. Chi-square greater). Each profile was analyzed separately from the others and therefore it represented a sub-sample, and we considered all wall posts, and the associated comments. We calculated $50$ values for the $Chi^2$ statistic.
The participants' profiles were assessed through the significant statistical of the parameter $Chi^2$. It was possible to identify the most empathetic profiles ($n = 8$), i.e. Facebook profiles for which a statistically significant relationship between post and comment mood was found ($Chi^2 > or = 4$). These participants were defined as subjects with highly empathetic profile. However, other profiles can be characterized also in terms of "intensity of the degree of empathy profile", on the basis of their $Chi^2$ absolute value associated with.

\subsection{Relation between Excessive Self-presentation and Empathy profile}

To evaluate the relationship between the Excessive Self-presentation style and the Empathy profile, first of all, a correlation analysis was carried out among the parameters used to estimate the empathy profile of each participant (i.e. $Chi^2$ value), with the Facebook variables, LIWC categories and Excessive Self-Presentation Model. The most significant results are the following.
Regarding the Facebook variables, the participants with more empathetic profiles publish a greater number of famous quotes ($r = .344, \textit{p.} < .05$), write longer average posts ($r = .288, \textit{p.} < .05$) and receive a greater number of comments ($r = .307, \textit{p.} < .05$).
About the LIWC categories, the participants with more empathetic profiles use more words relate to the physical ($r = .394, \textit{p.} < .01$), the body ($r = .352, \textit{p.} < .05$), the sensorial processes ($r = .305, \textit{p.} < .05$), more words associated with the possibility ($r = .309, \textit{p.} < .05$) and more commas ($r = .304, \textit{p.} < .05$). Finally, regarding the Excessive Self-presentation Model the participants with an excessive self-presentation elicit a higher number of empathetic comments by their Facebook friends compare to the other subjects  ($r = .273, \textit{p.} < .05$). Subsequently, a sample discretization was carried out based on the Empathetic detector representing the degree of coherence on the emotional profile analyzed (i.e. $Chi^2$ value associated with each Facebook Profile). We defined two sub-groups respectively called "Little empathetic Profiles" and "High Empathetic Profiles". The first group was composed by the Facebook Profiles that reported a lower $Chi^2$ score than the median of the entire sample ($median = .74$) while the second group was composed by Facebook profiles with higher scores. In table 4, the significant statistical tests describing the differences between the little and high empathetic profiles in relation to the followed variables are reported (e.g. Facebook variables, LIWC categories and the Excessive Self-Presentation Model). 

Student's t-test was conducted. About the Facebook variables, the participants with more empathetic profiles publish longer posts, more comments, more posts with comments, more personal photos, more posts, more average long posts, more complete activity, more activities with like, and more famous quotes. Concerning the LIWC categories, the participants with more empathetic profiles use more singular second person verbs, more words, more words related to physical, sensorial processes, possibility, sex, and money. These subjects utilize more present tense, singular first person pronouns, commas, conditional sentences, and less words with more than six letters. About the Excessive Self-presentation Model, the participants with more empathetic profiles report a higher score on our metric. (table \ref{tab:empathy_differences}) 

Moreover, the gender shows a significant effect, with the females appear as characterized by a more empathetic profile ($\textit{t}.=-3.324, p<.01$).

\begin{table*}[htbp!]
\caption[]{\textit{Differences between High and Low Empathetic Profiles}. The table reports the significant Student \textit{t} values describing the differences between users with high or low empathetic Facebook profiles. All the statistics reported in table are significant at a level of $\textit{p.} < 0.01$ (**) or $\textit{p.} < 0.05$  (*). The values regarding the LIWC variables are frequently represented by percentages.}
\label{tab:empathy_differences}
\begin{center}
\small
\begin{tabular}{lcccc}
\multicolumn{1}{c}{\bf Variable} &\multicolumn{1}{c}{\bf \textit{t}} &\multicolumn{1}{c}{\bf Group} &\multicolumn{1}{c}{\bf M} &\multicolumn{1}{c}{\bf SD} \\
 \hline
\multirow{2}{*}{Wall post length} & \multirow{2}{*}{-3.187**} &
Low & 30109 & 29312\\
&& High & 64910 & 46591\\
\multirow{2}{*}{Comments} & \multirow{2}{*}{-3.094**} &
Low & 882 & 518\\
&& High & 1630 & 1109\\
\multirow{2}{*}{Wall posts with comments} & \multirow{2}{*}{-2.953**} &
Low & 196 & 119\\
&& High & 335 & 206\\
\multirow{2}{*}{Personal photos} & \multirow{2}{*}{-2.885**} &
Low & 33 & 33\\
&& High & 66 & 47\\
\multirow{2}{*}{Wall posts} & \multirow{2}{*}{-2.809**} &
Low & 402 & 293\\
&& High & 767 & 589\\
\multirow{2}{*}{Wall post average length} & \multirow{2}{*}{-2.618**} &
Low & 71 & 24\\
&& High & 92 & 32\\
\multirow{2}{*}{Complete activity} & \multirow{2}{*}{-2.572*} &
Low & 472 & 352\\
&& High & 837 & 623\\
\multirow{2}{*}{Activities with like} & \multirow{2}{*}{-2.570*} &
Low & 468 & 348\\
&& High & 827 & 613\\
\multirow{2}{*}{Famous quotes} & \multirow{2}{*}{-2.217*} &
Low & 65 & 80\\
&& High & 125 & 110\\
\hline \
\multirow{2}{*}{Singular second person verb} & \multirow{2}{*}{-3.514**} &
Low & 1,42 & 0,63\\
&& High & 2,02 & 0,57\\
\multirow{2}{*}{Word Count} & \multirow{2}{*}{-3.301**} &
Low & 5138 & 5002\\
&& High & 11297 & 7966\\
\multirow{2}{*}{Physical} & \multirow{2}{*}{-3.082**} &
Low & 1,22 & 0,48\\
&& High & 1,59 & 0,35\\
\multirow{2}{*}{Sensorial processes} & \multirow{2}{*}{-2.954**} &
Low & 1,12 & 0,38\\
&& High & 1,37 & 0,17\\
\multirow{2}{*}{Possibility} & \multirow{2}{*}{-2.776**} &
Low & 1,54 & 0,49\\
&& High & 1,91 & 0,46\\
\multirow{2}{*}{Sexual} & \multirow{2}{*}{-2.628*} &
Low & 0,29 & 0,15\\
&& High & 0,40 & 0,14\\
\multirow{2}{*}{Present tense} & \multirow{2}{*}{-2.254*} &
Low & 7,98 & 1,59\\
&& High & 8,81 & 0,90\\
\multirow{2}{*}{Word with more than six letters} & \multirow{2}{*}{2.214*} &
Low & 20,95 & 2,53\\
&& High & 19,48 & 2,11\\
\multirow{2}{*}{Comma} & \multirow{2}{*}{-2.212*} &
Low & 3,29 & 1,50\\
&& High & 4,20 & 1,54\\
\multirow{2}{*}{Singular first person pronoun} & \multirow{2}{*}{-2.146*} &
Low & 1,88 & 0,64\\
&& High & 2,32 & 0,80\\
\multirow{2}{*}{Conditional} & \multirow{2}{*}{-2.092*} &
Low & 0,58 & 0,28\\
&& High & 0,74 & 0,23\\
\multirow{2}{*}{Money} & \multirow{2}{*}{-2.049*} &
Low & 0,09 & 0,06\\
&& High & 0,12 & 0,06\\
\hline \
\multirow{2}{*}{Excessive Self-presentation Model} & \multirow{2}{*}{-3.770**} &
Low & -0,39 & 0,61\\
&& High & 0,42 & 0,89\\
\hline \
\end{tabular}
\normalsize
\end{center}
\end{table*}

\section{Discussion and conclusions}

Confirming the literature \cite{Bazarova2012,Hu2015}, our results show that participants publish a higher average number of posts with positive emotional loading, compared to an average number of posts with negative emotional loading. Furthermore, it is confirmed that to publish negative emotional posts increases the likelihood of receiving negative comments \cite{Hancock2008,Guillory2011}. The research also highlights that to publish a post with emotional charge, either positive or negative, corresponds to get more likes, and comments, assuming a precise linguistic strategy for online self-presentation \cite{Bazarova2012}.
The general Chi-Square analysis showed a significant relationship between posts and comments mood, in particular there was a greater tendency to respond in a positive way, rather than negative, to positive posts and vice versa. The results appear to confirm the possibility of an emotional coherence through an Internet-based communication, stressing how emotional dissemination does not require face to face interaction, but can occur even during online interactions \cite{Kramer2012,Kramer2014}.
To compare the subjects with an \textit{High Empathetic profile} with those with a \textit{Low empathetic profile}, the sample has been split in two sub-samples, depending on the median reported to the Chi square test ($Median= 0.74$). Regarding the gender effect, previous studies have shown that females post more empathetic comments than males \cite{Preece2001}. Our study partially confirms such results, showing that the females profiles appeared to be more empathetic (i.e., a stronger emotional coherence between posts and comments mood) than those of males.
In the second part of our work we investigated the relations between the profile empathetic level, with the other factors of interest got into account in the present study. We found that the more empathetic profiles have a higher activity, receiving a higher number of likes for each activity. Such profiles are characterized by a greater number of wall posts, that tend to be longer and to receive more comments. The profiles with a greater empathy shown a greater number of personal photos. The linguistic and semantic analysis revealed how the more empathetic profiles use a higher number of singular first person pronouns, verbs in the singular second person, a greater number of words, and particularly more words related to sex and physical. Finally, the more empathetic profiles are characterized by a more excessive self-presentation style on SNSs. Such results seem to confirm the literature regarding the relationship between narcissism and excessive self-presentation on SNSs. Previous studies \cite{Buffardi2008,Ong2011} emphasized how the narcissistic trait is related to an intense activity on Facebook, such as connecting to many friends and publishing more wall posts and photos. In addition, another research adopting LIWC \cite{DeWall2011} found that narcissistic people published more self-promoting and sexy photos, and had a more aggressive language when singular first person pronouns were less employed. Moreover, \cite{Holtzman2010} pointed out that narcissistic people used to talk more about sexual topics, while \cite{Bergman2011} underlined how narcissists may be characterized by a high degree of personal information disclosure, status and personal photos updating. The main result of our study indicates how the subjects with an excessive self-presentation style elicit a greater density of empathetic comments. In general, the individuals reporting a greater score on the Excessive Self-presentation model are those who have the more empathetic and coherent emotional profiles. In this way, our study appears to confirm how to publish emotional contents on own profile can increase the likelihood of receiving more comments and likes, fulfilling a goal of attention-seeking, as narcissistic people have. Thus, to post on the profile positive and negative emotional contents might be considered an effective way to satisfy the that need. The study has two limitations: first, the socio-demographic features of participants who belong to the same municipality. To generalize and verify the two metric models, an extension toward different countries is needed. 
Also, an explicit measure to assess narcissism is missing. This is partially mitigated by using a real-world Facebook dataset and taking into account six validated narcissistic models to build our Index. Noteworthy, three of such models applied LIWC to evaluate the trait. Moreover, \cite{Deters2014} underlined the appropriateness of non-self-report measures to explore online behaviors because the linguistic analysis is objective and quantifiable behavioral data, and unlike surveys and questionnaires, it allows a “free” self-presentation in the users’ own words. Moreover, a self-report scale sometimes may encounter several difficulties (i.e. people not answering all questions, social desirability bias, etc.) \cite{Schwartz2013}. Nevertheless, a confirmatory study is recommended to verify the construct validity. Unfortunately, few studies investigated the empathetic coherence between posts and received comments, which precludes to fully compare these results on literature. Further studies are suggested, to increase knowledge on linguistic strategies of online self-presentation on SNSs. The main scientific contribution of our work is the processing of new effective measures for the assessment of emotional loading on SNSs posts, using little public information. Two ICT algorithms ("Positive Mood Indicator" and "Negative Mood Indicator") were applied to evaluate such a dimension. The LIWC analysis of the posts appears to be particularly recommended and suitable to increase the reliability of the measures, as \cite{Panek2013} suggests. The model, appearing robust even for short messages, could be used also for other SNSs, such as Twitter, Google+ both characterized by short messages, the typical style of the new web based virtual environments. Moreover, our study analyzed all posts on 50 profiles for one year, improving previous studies which attempted to merge posts contents and style \cite{Carr2012}.

\section*{Acknowledgments}
We thank the district of Prato for giving us the opportunity to collect data during the “ARCA Project”. We thank the high school "Gramsci-Keynes” in Prato for the support and availability for the data collection, as well as for the promotion of the research activities partially object of the present study.

%We wish to thank Franco Bagnoli, whose precious comments helped us in improving the contents and the quality of our paper, and Timoteo Carletti for the scientific contribution and support.

\bibliographystyle{elsarticle-num}
\bibliography{BIBLIO}

\end{document}